\documentstyle[epsfig]{mn}     % LaTeX A&A  Standard Fonts

\def\simless{\mathbin{\lower 3pt\hbox
     {$\rlap{\raise 5pt\hbox{$\char'074$}}\mathchar"7218$}}}   %< or of order
\def\simmore{\mathbin{\lower 3pt\hbox
     {$\rlap{\raise 5pt\hbox{$\char'076$}}\mathchar"7218$}}}   %> or of order
\def\msun{{\rm M}_\odot}                                       % solar masses
\newcommand{\lsi}    {LS~I +61$^{\circ}$ 235}

\begin{document}

   \title[Global oscillations in LS I +61$^{\circ}$ 235/RX J0146.9+6121]
   {Correlated V/R and IR photometric variations in the Be/X-ray binary LS 
I +61$^{\circ}$ 235/RX J0146.9+6121}

   \author[Reig et al.]
{P. Reig$^{1,2}$, 
I. Negueruela$^{3}$, M.J. Coe$^{4}$, J. Fabregat$^{5}$,
A.E. Tarasov$^{6}$ and R. K. Zamanov$^{7,8}$\\
$^{1}$ Foundation for Research and Technology-Hellas. 711 10 
Heraklion. Crete. Greece.\\
$^{2}$ Physics Department. University of Crete. 710 03 Heraklion. Crete. 
Greece.\\
$^{3}$ SAX SDC, Agenzia Spaziale Italiana, c/o Telespazio, via Corcolle 19, 00131 
Roma, Italy.\\
$^{4}$ Physics \& Astronomy Department, Southampton University, 
Southampton, SO17 1BJ, UK.\\
$^{5}$ Departament d'Astronomia i Astrof\'{\i}sica, Universitat de 
Val\'encia, E-46100 Burjassot-Val\'encia, Spain.\\
$^{6}$ Crimean Astrophysical Observatory, 334413 Nauchny, Crimea, 
Ukraine.\\
$^{7}$ National Astronomical Observatory Rozhen, P.O. Box 136, BG-4700 Smoljan, 
Bulgaria.\\
$^{8}$ Departamento de  F\'{\i}sica, Universidad de Ja\'en,
C/Virgen de la Cabeza 2, 23071 Ja\'en, Spain.
}

   \date{Received ????????; accepted ??????????}

   \maketitle

   \begin{abstract}
   
We report on the long-term variability of the Be/X-ray binary LS I
+61$^{\circ}$ 235/RX  J0146.9+6121. New  optical   spectroscopic  and 
infrared  photometric observations  confirm the presence of global 
one-arm   oscillations in the circumstellar disc of the Be star and allow
us to derive a V/R quasi-period of 1240$\pm$30 days. Pronounced shell events,
reminiscent of the spectacular variations in Be stars, are also seen. We
have found that the $J$, $H$ and $K$ infrared photometric bands vary  in 
correlation   with  the spectroscopic V/R variations, implying that the
one-armed disc oscillations are prograde. The effect of the  oscillations
is not only seen in the H$\alpha$  line but also in the He I
$\lambda6678$  and  Paschen  lines.  Since  these lines are formed at
different  radii in the equatorial  disc of the Be star, such effect
confirms the global nature of the  perturbation. The Keplerian disc has
been found to be denser than the average of a sample of isolated Be stars,
which may be indicative of some kind of interaction with the compact
companion. Finally, from a {\em Rossi X-ray Timing Explorer}  observation
we derive a spin period of the neutron star of 1404.5$\pm$0.5 s. 

\end{abstract}


\section{Introduction}

RX J0146.9+6121 is one of the slowest high mass X-ray pulsar systems
(Mereghetti et al. 1993; Hellier 1994). Its optical counterpart is the
$V=11.2$, B1III-V star  LS I +61$^{\circ}$ 235 located at an estimated
distance of 2.3$\pm$0.5 kpc (Motch et al. 1997; Coe et al. 1993; Reig  et
al. 1997a, hereafter R97). R97 derived the astrophysical  parameters of the
optical counterpart and  reported  V/R variations  with a quasi-period of
$\sim 3$ years. The line profile variability was attributed to the 
prograde  precession  of a one-armed  mode  confined in the Be star's
disc.  

Be/X-ray  binaries  comprise  approximately  70\% of the more  general
class of high mass X-ray  binaries,  the other  $\sim$~30\%  containing
evolved  (luminosity  class I and II) primaries.  In a Be/X-ray binary the
optical companion, a Be star, is characterised by an emission line
spectrum and an infrared  excess when compared to normal  B-type stars of
the same spectral  type. These two observational properties have their
origin in the cool gaseous quasi-Keplerian disc which lies on the equatorial
plane of the central star. A neutron star revolves  around the 10-20~$\msun$ 
primary in a rather  eccentric  orbit (e $\sim$  0.2--0.8) and
accretes material  expelled by the Be star from the disc -- in the form of a
low velocity, high density  equatorial wind -- giving rise to the X-rays. 

The physical properties of the Be star's disc in Be/X-ray binaries have
traditionally been considered to be the same as those of isolated Be
stars. Indeed, the long-term variability characterised as disc-loss phases
and V/R variations have been seen in both, isolated and Be/X-ray binaries
(Okazaki 1997; Negueruela et al. 1998).
However, it is not yet clear what role the compact companion in
Be/X-ray binaries may play in the onset and subsequent development of the
perturbation that gives rise to the asymmetric profiles or in the
formation and loss of the disc. There is growing evidence that the
circumstellar disc surrounding isolated Be stars and Be star in X-ray
binaries may not share, on average, the same physical properties (Reig et
al. 1997b; Negueruela et al. 1998; 1999)

In  this  paper  we  analyse  optical,  infrared  and  X-ray  data 
and search for correlations  between the  characteristics of the
radiation at  these  wavelengths. Our ultimate goal is to assess the
validity of the global one-armed oscillation model and whether the compact
companion has any effect on the V/R variability.

%----------------------------------------------------------------------------
\begin{table} 
\begin{center} 
\caption{Journal of spectroscopic observations and spectral parameters for
the H$\alpha$ line. Errors are $\simless$ 8\%}   
\label{sp} 
\begin{tabular}{lccccccl} 
\hline 
Date	&MJD	&Tele 		&Range  &EW	&$\Delta V$	&profile\\
	&	&scope		&nm     &\AA	&km/s		&\\ 
17-08-94& 49582 & BNAO$^a$	&650-662 &--8.4	 &252	&V$>$R  \\ 
08-06-96& 50243	& BNAO$^a$	&650-662 &--9.6  &257	&V$<$R  \\ 
09-06-96& 50244 & BNAO$^a$	&650-662 &--10.1 &258	&V$<$R  \\
01-08-96& 50297	& INT$^a$	&610-700 &--10.1&316	&V$<$R  \\ 
26-08-96& 50322	& BNAO$^a$	&650-662 &--10.2 &298	&V$<$R  \\ 
13-11-96& 50401	& CAO   	&653-660 &--10.3	&259	&V$<$R  \\ 
12-12-96& 50430 & CAO		&653-660 &--10.3	&251	&V$<$R  \\ 
01-02-97& 50481	& INT$^a$	&580-740 &--11.9&288	&V$<$R  \\
21-08-97& 50682	& INT$^a$	&610-700 &--12.1&242	&V$>$R  \\ 
18-09-97& 50710	& CAO		&653-660 &--12.0	&223	&V$>$R  \\ 
10-10-97& 50732 & BNAO$^a$	&650-662 &--10.7	&233	&V$>$R  \\ 
28-10-97& 50750	& JKT		&638-673 &--11.8	&258	&V$>$R  \\ 
10-02-98& 50855	& INT   	&640-680 &--11.2&235	&V$>$R  \\
01-10-98& 51088	& BNAO$^b$	&650-670 &--11.0	&337	&V$>$R  \\ 
04-10-98& 51091 & BNAO$^b$	&650-670 &--11.3 &333	&V$>$R  \\ 
08-10-98& 51095 & WHT   	&649-661 &--11.5	&336	&V$>$R  \\ 
22-07-99& 51382 & CAO   	&653-660 &--10.8 &298	&shell  \\
25-07-99& 51385 & SKI   	&555-756 &--11.9 &312	&shell  \\ 
26-07-99& 51386 & SKI   	&555-756 &--11.6 &325	&shell  \\ 
19-08-99& 51410 & INT$^b$	&624-680 &--11.2&298	&shell  \\ 
17-09-99& 51439 & BNAO$^b$  	&650-670 &--12.2 &286	&shell  \\
19-09-99& 51441 & BNAO$^b$  	&650-670 &--11.7 &285	&shell  \\ 
\hline
\multicolumn{5}{l}{BNAO$^a$: ISTA 580$\times$400 + 632 l/mm} & \multicolumn{2}{l}{0.2 \AA/pix}  \\
\multicolumn{5}{l}{BNAO$^b$: Photometrics 1024$\times$1024 + 632 l/mm} & \multicolumn{2}{l}{0.2 \AA/pix} \\
\multicolumn{5}{l}{CAO: Electronix 1024$\times$260 CCD + 600 l/mm} &\multicolumn{2}{l}{0.06 \AA/pix} \\ 
\multicolumn{5}{l}{JKT: TEK4 + RBS + 2400 l/mm}  &\multicolumn{2}{l}{0.4 \AA/pix} \\ 
\multicolumn{5}{l}{INT$^a$: TEK3 + IDS + 235 mm + R1200 l/mm} & \multicolumn{2}{l}{0.8 \AA/pix} \\ 
\multicolumn{5}{l}{INT$^b$: EEV10 + IDS + 500 mm + R1200 l/mm} &\multicolumn{2}{l}{0.2 \AA/pix} \\ 
\multicolumn{5}{l}{WHT:	UES + 31 l/mm, R=54000} &\multicolumn{2}{l}{0.06 \AA/pix}\\ 
\multicolumn{5}{l}{SKI: ISA608 2000$\times$800 + 1300 l/mm} & \multicolumn{2}{l}{1 \AA/pix}
\end{tabular} 
\end{center} 
\end{table}
%----------------------------------------------------------------------------
%----------------------------------------------------------------------------
\begin{table} 
\begin{center} 
\caption{Journal of infrared observations}  
\label{irobs} 
\begin{tabular}{cccccccc} 
\hline 
Date	&MJD	&J	&H	&K \\ 
13-01-96	&50096.42    &9.86$\pm$0.03   &9.58$\pm$0.02   &9.35$\pm$0.02\\ 
19-03-96	&50162.36    &9.78$\pm$0.02   &9.60$\pm$0.01   &9.40$\pm$0.02\\ 
28-07-96	&50293.68    &9.73$\pm$0.05   &9.51$\pm$0.06   &9.34$\pm$0.04\\
29-07-96	&50294.65    &9.73$\pm$0.10   &9.51$\pm$0.09   &9.31$\pm$0.08\\ 
18-07-97	&50648.65    &9.83$\pm$0.02   &9.56$\pm$0.02   &9.32$\pm$0.02\\ 
19-07-97	&50649.66    &9.77$\pm$0.01   &9.52$\pm$0.01   &9.32$\pm$0.01\\ 
21-07-97	&50651.66    &9.76$\pm$0.02   &9.53$\pm$0.02   &9.29$\pm$0.02\\
17-06-98	&50982.72   &10.01$\pm$0.01   &9.81$\pm$0.02   &9.53$\pm$0.02\\ 
27-10-98	&51114.42   &10.01$\pm$0.02   &9.69$\pm$0.02   &9.48$\pm$0.02\\ 
28-10-98	&51115.67    &9.82$\pm$0.02   &9.66$\pm$0.02   &9.45$\pm$0.02\\ 
28-10-98	&51115.67    &9.85$\pm$0.02   &9.65$\pm$0.02   &9.44$\pm$0.02\\
26-07-99	&51386.72    &9.80$\pm$0.02   &9.58$\pm$0.02   &9.37$\pm$0.02\\ 
27-07-99	&51387.66    &9.96$\pm$0.03   &9.67$\pm$0.02   &9.43$\pm$0.02\\ 
29-07-99	&51389.63    &9.85$\pm$0.02   &9.61$\pm$0.02   &9.36$\pm$0.02\\ 
30-07-99	&51390.66    &9.81$\pm$0.02   &9.58$\pm$0.02   &9.34$\pm$0.02\\
01-08-99	&51392.64    &9.90$\pm$0.04   &9.63$\pm$0.03   &9.41$\pm$0.02\\ 
02-10-99	&51454.49    &9.79$\pm$0.02   &9.54$\pm$0.02   &9.33$\pm$0.02\\ 
03-10-99	&51455.60    &9.86$\pm$0.02   &9.58$\pm$0.02   &9.32$\pm$0.02\\ 
04-10-99	&51456.56    &9.82$\pm$0.02   &9.57$\pm$0.02   &9.31$\pm$0.02\\ 
\hline
\end{tabular} 
\end{center} 
\end{table}
%----------------------------------------------------------------------------
%----------------------------------------------------------------------------
        \begin{figure*}
    \begin{center}
    \leavevmode
\epsfig{file=figures/serie.eps, width=14.0cm, bbllx=20pt, bblly=315pt,
  bburx=535pt, bbury=710pt, clip=}
 \end{center}              
        \caption{Evolution of the H$\alpha$ line profile over the 
period 1991 August 28 -- 1999 September 17 (MJD 48497--51439)}
        \label{vr}
        \end{figure*}
%----------------------------------------------------------------------------

\section{Observations and results}

\subsection{Optical data}

Optical  spectroscopic  observations were made from the 2.6m telescope at 
the  Crimean   Astronomical   Observatory   (CAO),  the 2m RCC telescope
of the Bulgarian National Astronomical Observatory "Rozhen" (BNAO), the
1.3m telescope of the Skinakas Observatory (SKI), in Crete (Greece), the
1.0m Jacobus Kapteyn  Telescope (JKT), the 2.5m Isaac Newton  Telescope 
(INT) in service mode and the 4.2m William Herschel  Telescope (WHT). The
last three telescope are located in the Observatorio del Roque de Los
Muchachos (La Palma, Spain). Table~1 shows the journal of the observations
and the instrument set-up. For  a description of the observations not
mentioned in the table see R97.

Fig \ref{vr} shows a selected sample of H$\alpha$ line profiles of LS I
+61$^{\circ}$ 235 covering the period 1991 August -- 1999 September. Part
of the new data (all  observations  taken after 1996  February  plus those
from Rozhen BNAO) are plotted together with some of the spectra presented
in R97. The evolution of the V/R ratio can be seen in Fig \ref{corr}$a$.
Different shapes of the line have been represented by different symbols.
Stars are used for $ V > R$ points, triangles for $V < R$ phases, dots for
single-peak lines and squares for shell profiles (i.e. when the central
absorption goes beyond the continuum). Note that the shell phase is very
brief and always occurs during the transition from $V > R$ to $V < R$. In
contrast, the transition from $V < R$ to $V > R$ is separated by
single-peak profiles. Strictly speaking there are not symmetric
single-peak lines in our observations (as observed in pole-on stars) since
one can always see flank inflections revealing a second peak.
Nevertheless, we will refer to as single-peak phase the profiles seen
during the transition  $V < R$ to $V > R$. By fitting a sine  function to
the V/R curve we refined the V/R quasi-period to a value of 1240$\pm$30
days.  

We have also measured the separation of the blue and red peaks by fitting
two Gaussian functions to the line profile (Fig~\ref{corr}$d$). The peak
separation gives a measure of the velocity field, assuming it to be
Keplerian. The blue-dominated profiles seem to sample a wider range of
velocities than red-dominated profiles: $V > R$ points spread over a
velocity range 220--360 km s$^{-1}$, whereas $V < R$ points distribute
around 250--320 km s$^{-1}$. Also, blue-dominated profiles with high
values of the peak separation occur only at the end of the $V > R$ cycle,
just before the shell phase. 

The density wave does not only affect the H$\alpha$  line but also the
Paschen series and He I $\lambda$6678 (Fig~\ref{global}). Since  these 
lines  are  formed  at  different   regions  inside  the equatorial disc
the perturbation  must extend over a very wide region, hence  confirming
its global  nature. 

%----------------------------------------------------------------------------
        \begin{figure}
    \begin{center}
    \leavevmode
\epsfig{file=figures/corr2.eps, width=7.0cm, bbllx=48pt, bblly=156pt,
  bburx=293pt, bbury=720pt, clip=}
 \end{center}             
	\caption{Evolution of the V/R ratio ({\em a}), J band ({\em b}),
K band ({\em c}), J-K index ({\em d}), 
H$\alpha$ equivalent width ({\em e}) and peak separation ({\em f}).
A correlation between the V/R cycle and the infrared emission is apparent. 
The arrow in panel {\em (e)} indicates the X-ray outburst}
        \label{corr}
        \end{figure}
%----------------------------------------------------------------------------

\subsection{Infrared data}

The  infrared  observations  were taken with the  Continuous  Variable
Filter (CVF)  infrared  photometer,  using the 1.5-m Carlos  S\'anchez
telescope,  located at the Teide  Observatory,  in  Tenerife  (Spain). The
data were reduced following the procedure described by Manfroid (1993).
Instrumental values were transformed to the TCS standard system (Alonso et
al. 1998). Table~\ref{irobs} shows the results of the infrared
observations. For earlier observations the reader is referred to R97.

The  combination  of the old  and  new  infrared  data  yields  a very
interesting result, namely the correlation  between the V/R variations and
the infrared  magnitudes.  Fig~\ref{corr} shows the J and K light curves and
the evolution of the colour index J-K, covering the
period 1991 August to 1999 October.  The light curves were rebinned into 30
day bins. The errors were calculated from the photometric errors depending
on the number of points, $N$, in each bin: when the corresponding bin
contained only one point we simply took the observation error, if $1 < N
\leq 5$ then we defined the error as $\sum|x_i-x_m|/N$ and if $N > 5$ then
we considered the standard deviation  $\sqrt{(x_i-x_m)^2/N}$. 

As can be seen in Fig \ref{corr}, there is a distinct  modulation with an
amplitude, from maximum to minimum, of $\sim$ 0.3 mag.  The period of
this  modulation is  3.1--3.8  years, in  good  agreement  with the 
optical  V/R variability of 3.4 years. Interestingly, the infrared maxima
occur in coincidence with the optical ($\sim$ MJD 49250) and X-ray ($\sim$
MJD 50630) outbursts. On  the  other   hand,   the   infrared   colours 
do  not  change  drastically  over  the  period  of  the observations. 
That is, the slope of the infrared continuum remained the same over the
time covered by the observations.

\subsection{X-ray data}

LS I +61$^{\circ}$ 235 is  a persistent low-luminosity   Be/X-ray  
binary.  These systems   are   characterised   by long pulse periods, low
X-ray variability  ($L_{max}/L_{min}  \simless 10$) and low but 
permanent  levels of X-ray  emission  (Reig \& Roche 1999).  

RX J0146.9+6121 was observed with the RXTE {\em Proportional Counter
Array} (PCA) on March 21, 1998 (02:01--11:54 UT). Good time  intervals 
were defined by removing data taken at low Earth elevation angle ($<$ 
10$^{\circ}$)  and during times of high particle background.  An  offset 
of only  0.02$^{\circ}$  between  the  source position and the pointing of
the satellite was allowed, to ensure that any possible short stretch of
slew data at the beginning and/or end of the  observation  was  removed. 
All five PCA units  were  functioning during the entire  observation.  The
total net exposure was 19223 s. Due to the relatively wide field of view
of the PCA instrument (1$^{\circ}$ FWHM) the nearby X-ray source 4U
0142+61 (White  et al.  1996) also contributed to the total flux. After 
correcting for collimator  efficiency the contribution of 4U 0142+61 to
the total observed  flux in the energy  range  2--30 keV was estimated  to
be of $\sim$ 10\%.  Because of this no X-ray spectral analysis was
attempted.

The pulse period was  determined by  correcting  the data to the
solar system  barycentre  and using the epoch  folding  technique, i.e, we
folded the data over a range of periods  and  searching  for a maximum 
$\chi^2$ as a function of period.  The pulse period found was
1404.5$\pm$0.5 s, which is virtually the same as the  1404.2$\pm$1.2 s
pulse period obtained in another RXTE  observation  nine months earlier
(Haberl et al. 1998). 
 
%----------------------------------------------------------------------------
        \begin{figure}
    \begin{center}
    \leavevmode
\epsfig{file=figures/global.eps, width=8.0cm, bbllx=20pt, bblly=310pt,
  bburx=488pt, bbury=712pt, clip=}
 \end{center}              
	\caption{The V/R oscillations are not only seen in H$\alpha$ ({\em
left}) but also in He~I $\lambda6678$ ({\em middle}) and Paschen lines Pa11
($\lambda8863$) and Pa12 ($\lambda$8750) ({\em right})}
        \label{global}
        \end{figure}
%---------------------------------------------------------------------------

\section{Discussion}
\subsection{Global $m=1$ oscillations}

In R97 we  investigated  the  different  models  that had been put forward
to explain the V/R  variability in Be stars and concluded that the model
which best accounted for the  observational  data in LS I +61$^{\circ}$ 235
was the Global  One-armed  Oscillation  model (Okazaki 1991, 1997;
Papaloizou et al. 1992).  This model  suggests that the  long-term  V/R 
variations  are caused by global $m=1$ oscillations in the cool equatorial
disc of the Be star.  In other words, an enhanced density perturbation
develops on one side of the disc, which  slowly  precesses.  The 
precession  time being that associated with the V/R  quasi-period.  The
density  perturbation is confined  within a few stellar radii in the disc
and the precession period turns out to be fairly insensitive to the size
of the disc (Savonije \& Heemskerk 1993).

One  prediction  of the model is that no  changes  in the slope of the
infrared   continuum  are   expected.  The  reason  is  that  the  V/R
variations are not the result of changes in the radial gradient of the
circumstellar  gas.  The slope of the infrared  continuum is a measure of
the radial density distribution but since matter in the disc does not
move  radially  no changes in the shape of the  infrared continuum are
expected.  This is exactly the behaviour that we find in the case of LS
I +61$^{\circ}$ 235.  While the  individual  infrared  photometric bands
changed  ($\Delta J \approx \Delta H \approx  \Delta K \sim 0.3$
magnitudes) the infrared colours remained unchanged (Fig.~\ref{corr}).

In  principle,  the issue of whether  the  motion of the  perturbation
occurs  in the  same  sense  (prograde  rotation)  or  opposite  sense
(retrograde  rotation) to the stellar rotation can be found out from the 
observations.  Telting et al.  (1994) realised that a prograde revolution
implies that the $V > R$ phase must be followed by a shell profile  and a 
similar profile but with much less pronounced absorption feature (or
possibly a single peak line if the inclination is low) during the
transition from $V < R$ to $V > R$.  These characteristic  line shapes
must translate into noticeable  photometric variations.  According  to 
Mennickent et al. (1997),  we should expect a minimum of brightness when
$V=R$ {\em prior} to the $V < R$ phase if the  motion  is  prograde  and
$V > R$  {\em after}  $V=R$  if retrograde.  In LS I+61 235 the minimum of
brightness  in the infrared photometric  bands occurred  during the shell
phase ($V=R$) before the $V < R$ phase began,  confirming the prograde
nature of the precession inside the disc.

However, models of one-armed global density waves  cannot reproduce 
strong  shell-non-shell transitions like the ones seen in \lsi\, which are
reminiscent of the so-called {\em spectacular variations} (Doazan et al.
1983). Such shell events seem to be rare a phenomenon which have been
reported for only three Be stars: $\gamma$ Cas, 59 Cyg and Pleione (Hummel
1998 and references therein), all of which are either binaries or 
suspected binaries. One possible explanation might be a {\em thick} disc in the
region where the perturbation lies.  When the perturbation is behind the
star and for the right inclination angle, no shell event is seen because
the disc in between the central star and the observer is thin and does not
occult the star nor the perturbation. The shell phase would occur
when the perturbation is at inferior conjunction since the thicker disc
would hide the central star from the observer. An alternative explanation
is given by Hummel (1998) who suggested a tilted or warped circumstellar
disc with precessing nodal line in addition to the density wave (see also
Porter 1998).

\subsection{A dense circumstellar disc in \lsi}

The works by Dachs et al. (1986) and Hanuschik et al. (1988) have shown
that the equivalent width of H$\alpha$ line emission for Be stars
increases with the effective disc radius.  Since for rotationally
dominated  profiles $\Delta V/(2\,v\,\sin{i})\;$ can be regarded as a
measure  of the radius of the H$\alpha$ emitting region (Huang 1972), we
expect a correlation between the peak separation and the H$\alpha$
equivalent width. Hanuschik et al. (1988) derived the following law

\begin{equation}
\log \left(\frac{\Delta V}{2 v \sin i} \right)= a \log (EW(H\alpha)) + b
\end{equation}

\noindent where $v \sin i$ is the projected rotational velocity and
EW(H$\alpha$) is given in anstrongs. $a$ and $b$  are related to the
rotational law index $a=-j/2$ ($j$=0.5 for Keplerian rotation and $j$=1
for conservation of angular momentum) and with the disc electron density,
respectively. A least square fit to the LS I +61$^{\circ}$ 235 data gave
$a=-0.23\pm0.10$, i.e. $j\approx 0.5$ and $b=+0.1\pm0.1$. These values are
to be compared with the average values $a=-0.4$ and $b=-0.1$ found by
Hanuschik et al. (1988) for a sample of 26 isolated Be stars.  The higher
value of $b$ in \lsi\ implies a denser disc than those of isolated Be
stars. 

The main (and probably only) difference between a Be/X-ray binary and an
early-type isolated Be star is the presence of a neutron star in the
former. It then seems natural to attribute the dissimilarity in the
properties of the circumstellar envelopes to the influence of such compact
companion. The neutron star trims the disc to a certain radius and
prevents its free growth, making it denser. Disc truncation has been
suggested in other Be/X-ray binaries like V0332+53 (Negueruela et al.
1999). This idea  would support the hypothesis that the neutron star 
plays a fundamental role in the evolution and properties of the equatorial
disc in Be/X-ray binaries, as proposed by Reig et al. (1997b).

\subsection{X-ray/optical/IR correlations}

In R97  the  observation  of an optical  outburst around 1993 
September-October (MJD $\sim$ 49260) was reported. During the outburst the
H$\alpha$ equivalent width, EW(H$\alpha$),  changed by $\sim$ 10 \AA, in
about 270 days, decreasing to  pre-outburst  values  ($\sim  -8 $\AA) in
about the same amount of time (Fig \ref{corr}$c$).  This increased 
coincided with a single-peak  phase of the H$\alpha$  profile.  The
question of whether this  outburst was an  isolated  event  or was
associated with the V/R cycle remained open due to the short coverage of
the data.  The new observations show no new EW(H$\alpha$)  maximum.  After
the  outburst  the  EW(H$\alpha$)  increased  slowly  up to a level of
$-12$ \AA, considerably lower than the peak of 1993 September.  The new
single-peak  phase should have occurred during 1997 March-May. 
Unfortunately,  the star was too close to the Sun to be observed. 
However, we notice that EW(H$\alpha$) seems to have reached a maximum
value just before and after the period  when  we  would  expect  the 
single-peak   phase  (around  MJD 50500). Thus, we are inclined   to think
that the higher EW(H$\alpha$) associated with single-peak profiles
reported in R97 may   reflect  the  fact  that the  fluxes  of both
components are adding up and are not affected by the  absorption feature 
present  in the other  profiles.  In this  context  then, the increase in 
EW(H$\alpha$) is real and  would be an event related to the motion of the
density pattern in the disc, rather than an isolated episode.

In 1997 July (MJD 50634) \lsi\ underwent a small X-ray outburst 
(Haberl et al. 1998).  The X-ray  luminosity increased by
nearly a factor of five in one week reaching  3.45  $\times$ 10$^{35}$ erg
s$^{-1}$ in the energy range 0.5--10 keV. This outburst is marked with and
arrow in Fig~\ref{corr}$c$. Interestingly, the outburst occurred at the
time of the expected infrared and EW(H$\alpha$) maxima.   One is then
tempted to attribute this correlated X-ray/optical/IR behaviour to the
high density perturbation where most of the Balmer emission is formed. If
the inclination of the system is less than 90$^{\circ}$,  when the
high-density part of the equatorial disc is behind the star it offers the
largest geometric area (especially so if the disc is thicker in this
region) and the highest optical and infrared emission. If the neutron star
happens to be close to the Be star it will accrete from this high-density
material and the X-ray emission will be enhanced. New optical and X-ray
observations around the next expected maximum are needed to solve this
issue.

\section{Conclusion}

Optical  spectroscopic  observations confirm  the  presence  of  global  
one-armed oscillations  in  the circumstellar  disc  of \lsi.  These 
oscillations  manifest themselves as quasi-periodic  variations in the
shape of the H$\alpha$ line, whose  asymmetric  double peak profile 
alternates  between red- (V$<$R) and  blue-dominated  (V$>$R)  emission.
The system also goes through Be-Be shell transitions, which might indicate
an asymmetric vertical structure of the disc in the form of a thick  or a
tilted disc. The V/R quasi-period is determined to be $\sim$ 1240$\pm$30
days. We have found a correlation between the infrared emission and the
V/R variations.  This is the first time that such  correlation is reported
in a Be/X-ray  binary. From the pattern traced by the IR light curves in
relation to the V/R ratio we conclude that the one-armed disk oscillations
are prograde. The Be star's disc in \lsi\ is found to be denser than 
isolated Be stars, which may be connected to the presence of the neutron
star.  From a {\em Rossi X-ray Timing Explorer}  observation we derive a
spin period of the neutron star of 1404.5$\pm$0.5 s. 

%The V/R variability is not exclusive of the H$\alpha$  line but is also
%present in the He I  $\lambda6678$  and  Paschen lines.  Since  these 
%lines  are  believed  to form at very  different depths in the Be star's 
%circumstellar  envelope we conclude  that the oscillations are affecting a
%large portion of the circumstellar  disc, i.e.  they are indeed global.
%The Be star's disc in \lsi\ is found to be denser than  isolated Be
%stars.  From a {\em Rossi X-ray Timing Explorer}  observation we derive a
%spin period of the neutron star of 1404.5$\pm$0.5 s. 

\subsection*{Acknowledgements}

We thank Chris  Moran for  providing  us with one of the INT  spectra and
Dr E. V. Paleologou for helping us with the spectroscopic observations at
Skinakas Observatory. Skinakas Observatory is a collaborative project of
the University of Crete, the Foundation for Research and Technology-Hellas
and the Max-Planck-Institut für Extraterrestrische Physik. P.  Reig 
acknowledges  support  via  the  European  Union Training and Mobility of
Researchers Network Grant  ERBFMRX/CT98/0195. IN is supported by an ESA
external fellowship. R. Zamanov acknowledges support from  Direcci\'on
general de relaciones culturales y  cient\'{\i}ficas, Spain. Some
observations were taken as part of the ING service observing programme. We
are grateful to the referees, Dr D. Baade and Dr T. Rivinius, for useful
comments.


\begin{thebibliography}{}
\bibitem{} Alonso A., Arribas S., Mart\'{\i}nez-Roger C. 1998, A\&AS 131, 209
\bibitem{} Coe M.J., Everall C., Norton A.J., Roche P., Unger S.J., Fabregat J., 
Reglero V. \& Grunsfeld J.M., 1993, MNRAS, 261, 599
\bibitem{} Dachs, J., Hanuschik, R., Kaiser, D. \& Rohe, D. 1986, A\&A, 159,
276
\bibitem{} Doazan V., Franco M., Rusconi L., Sedmark G., Stalio R., 1983,
A\&A, 128, 171
\bibitem{} Haberl F., Angelini L. \& Motch C., 1998, A\&A, 335, 587
\bibitem{} Hanuschik R.W., Kozok J.R., Kaiser D., 1988, A\&A 189, 147
\bibitem{} Hellier C., 1994, MNRAS, 271, L21
\bibitem{} Huang, S. 1972, ApJ, 171, 549
\bibitem{} Hummel W., 1998, A\&A 330, 243
%\bibitem{} Hummel W. \& Vrancken M., 1995, A\&A, 302, 751.
%\bibitem{} Hummel W., Hanuschik R.W, Vrancken M., 1998, in Kaper L., Fullerton 
%A.W., eds., Cyclical Variability in Stellar Winds, Proceedings of the ESO 
%Workshop held at Garching, Germany, 14-17 October 19. Springer-Verlag, 1998, 
%Berlin, p.343
%\bibitem{} Hummel W., Hanuschik R.W., 1997, A\&A 320, 852
%\bibitem{} Israel G.L., Mereghetti S. \& Stella L., ApJ, 1994, 433, L25
\bibitem{} Manfroid J. 1993, A\&A 271, 714
\bibitem{} Mennickent, R. E., Sterken, C. \& Vogt, N. 1997, A\&A, 326, 1167
\bibitem{} Mereghetti S., Stella L. \& De Nile F., 1993, A\&A, 278, L23
\bibitem{} Motch C., Haberl F., Dennerl K., Pakull M.\& Janot-Pacheco E.,
1997, A\&A, 323, 853
\bibitem{} Negueruela I., Reig P., Coe M.J. \& Fabregat J., 1998, A\&A,
336, 251
\bibitem{} Negueruela I., Roche P., Fabregat J., Coe, M. J., 1999, MNRAS,
307, 695
%\bibitem{} Negueruela I. \& Okazaki A.T., 1999, A\&A, in press
\bibitem{} Okazaki A.T., 1991, PASJ 43, 75
\bibitem{} Okazaki A.T., 1997, A\&A 318, 548
\bibitem{} Papaloizou, J.C., Savonije, G. J., Henrichs, H. F, 1992, A\&A, 265, 
L45
\bibitem{} Porter J.M., 1998, A\&A 336, 966
\bibitem{} Reig P., Fabregat J., Coe M. J., Roche P., Chakrabarty D., Negueruela 
I \& Steele I., 1997a, A\&A, 322, 183, (R97).
\bibitem{} Reig P., Fabregat J. \& Coe M.J., 1997b, A\&A, 322, 193
\bibitem{} Reig P. \& Roche P., 1999, MNRAS, 306, 100
\bibitem{} Savonije G.J. \& Heemskerk M.H.M., 1993, A\&A, 276, 409
%\bibitem{} Stella L., Israel G.L. \& Mereghetti S., 1998, in{\em The Many Faces 
%of Neutron Stars}, Kluwer Academic Publishers. Ed Buccheri et al.
\bibitem{} Telting, J.H., Heemskerk M.H.M., Henrichs H.F. \& Savonije G.J.,
1994, A\&A, 288, 558
\bibitem{} White N.E., Angelini, L., Ebisawa K., Tanaka Y. \& Ghosh P., 1996, 
ApJ, 463, L83

\end{thebibliography}
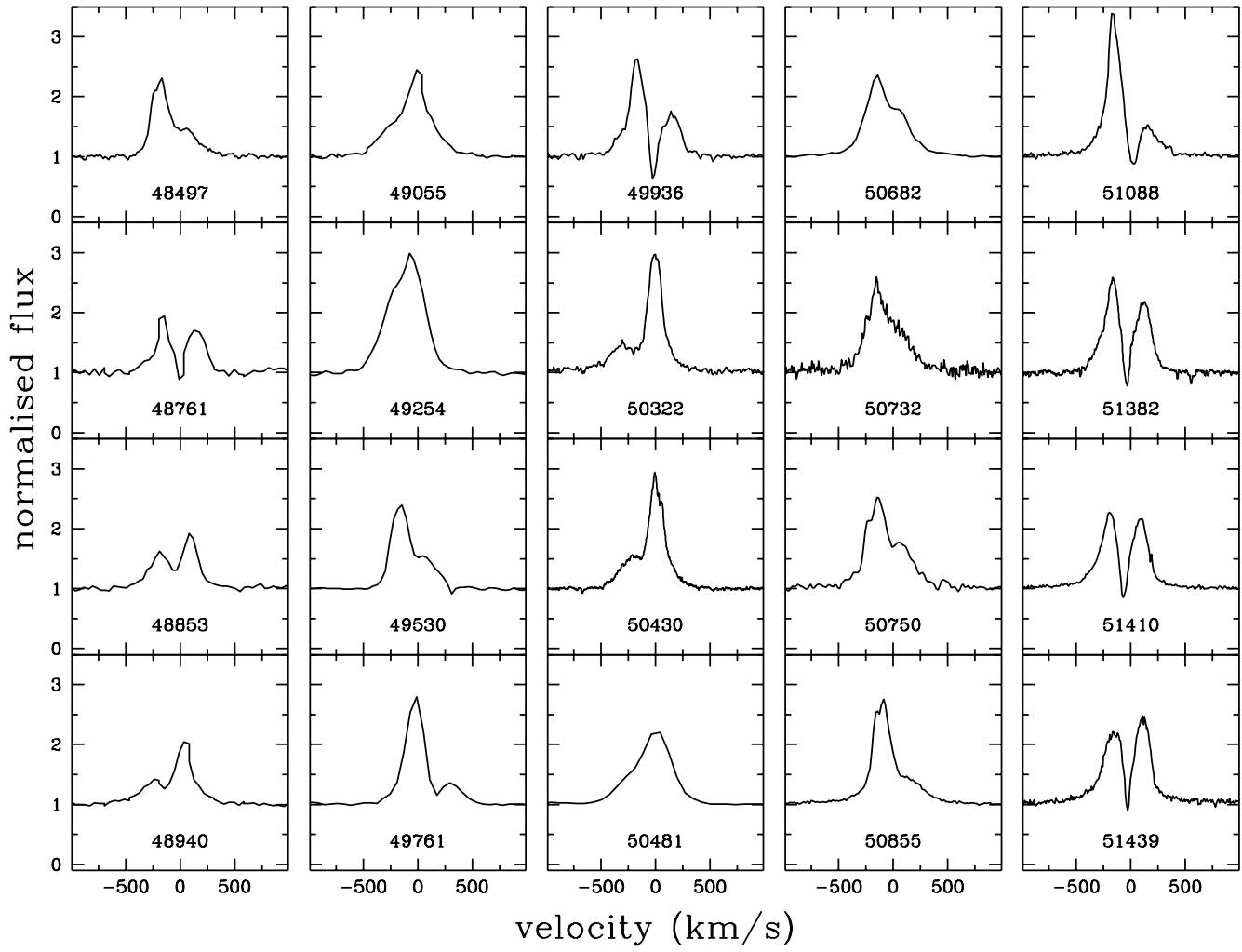
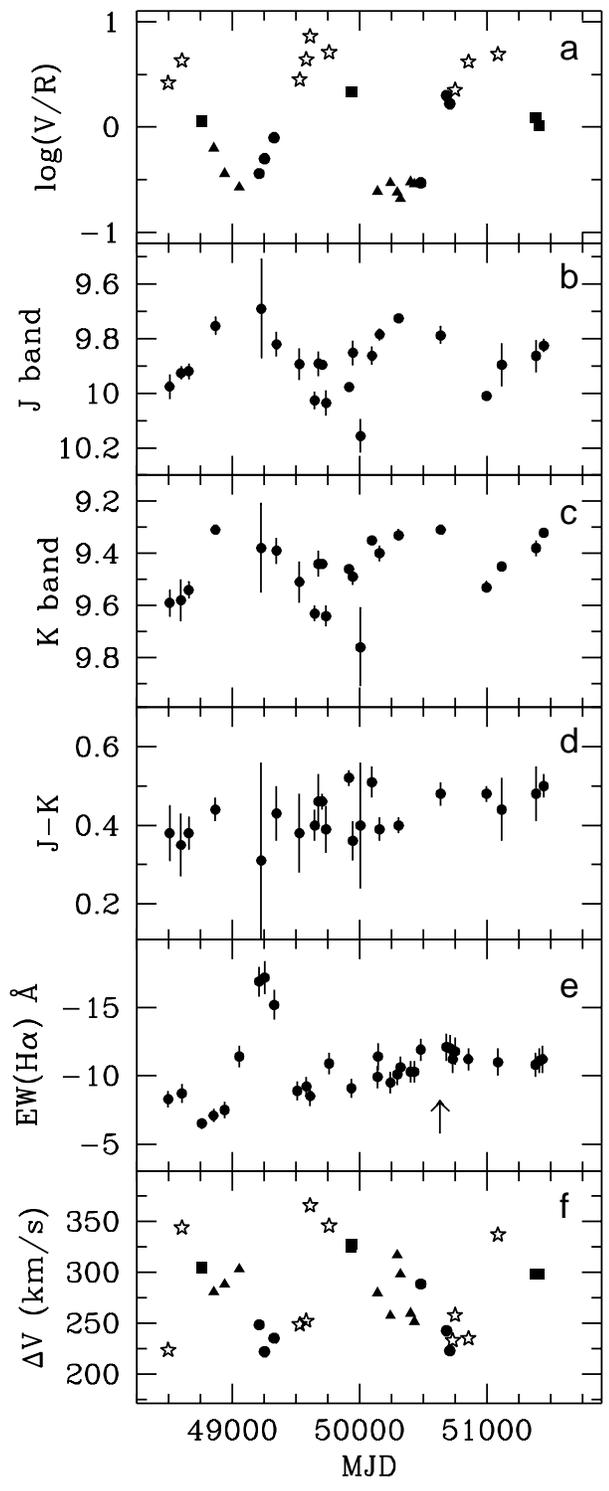
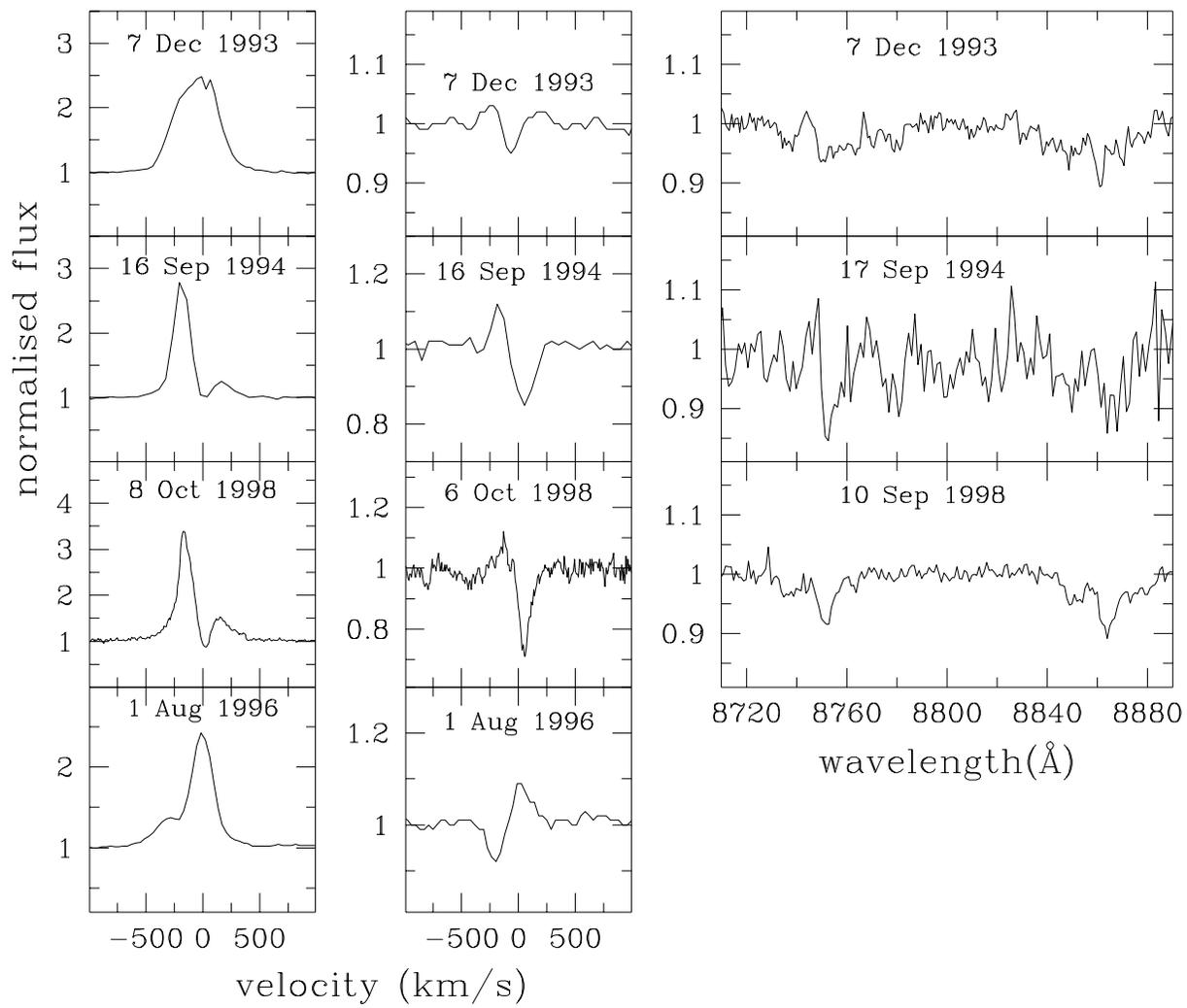
\end{document}